# Peculiarities of neutron waveguides with thin Gd layer

Yu Khaydukov[1,2,3], E Kravtsov[4,5], V Progliado[4], V Ustinov[4], Yu Nikitenko[6], T Keller[1,2], V Aksenov[7], and B Keimer[1]

[1] Max-Planck-Institut für Festkörperforschung, D 70569 Stuttgart, Germany
[2] Max Planck Society Outstation at the FRM-II, D-85747 Garching, Germany
[3] Skobeltsyn Institute of Nuclear Physics, 119991 Moscow State University, Moscow, Russia
[4] Institute of Metal Physics, 620990 Ekaterinburg, Russia
[5] Ural Federal University, 620002 Ekaterinburg, Russia
[6] Joint Institute for Nuclear Research, 141980 Dubna, Russia
[7] Petersburg Nuclear Physics Institute, 188300 Gatchina, Russia

E-mail: y.khaydukov@fkf.mpg.de

**Abstract**. Peculiarities of the formation of a neutron enhanced standing wave in the structure with a thin highly absorbing layer of gadolinium are considered in the article. An analogue of the *poisoning effect* well known in reactor physics was found. The effect is stronger for the Nb/Gd/Nb system. Despite of this effect, for a Nb/Gd bilayer and a Nb/Gd/Nb trilayer placed between $Al_2O_3$ substrate and Cu layer, it is shown theoretically and experimentally that one order of magnitude enhancement of neutron density is possible in the vicinity of the Gd layer. This enhancement makes it possible to study domain formation in the Gd layer under transition of the Nb layer(s) into the superconducting state (cryptoferromagnetic phase).

## 1. Introduction

Gadolinium due to the presence of a unique set of magnetic and nuclear properties is widely used in basic research and industrial application (electronics, nuclear energetic, medical and laser technique etc). One further possibly interesting application of Gd is cryptoferromagnetism (CFM) [1]-[3]. A sufficiently thin layer of a weak ferromagnet (F) being in contact with a superconductor (S) is predicted to go into a domain state below the superconducting transition temperature $T_C$. The domain size in CFM phase is comparable to the superconducting correlation length $\xi_S$, so that the effective exchange field of the ferromagnet on the size of the Cooper pair is close to zero, thereby providing the coexistence of superconducting and ferromagnetic orders. A phase transition into the CFM phase is possible in ferromagnets with the Curie temperature $T_m$ not much exceeding the superconducting transition temperature $T_C$. Gadolinium as F layer being in contact with superconducting niobium is predicted to be one of the best candidates for observing this effect [3]. The Curie temperature of bulk Gd is $T_m = 293K$. However, it can be suppressed in thin films down to 20-40K which is comparable with $T_C \sim 10K$ of Nb layer. Moreover, Gd and Nb are a well known pair of metals which are not solvable, allowing thus creating a sharp Gd/Nb interface [4]. Another distinctive feature of the Gd/Nb systems prepared by magnetron sputtering is high transparency of the S/F interface for Cooper pairs [4],[5].

The most direct method for detecting the CFM phase transition is Grazing Incidence Small Angle Neutron Scattering (GISANS, see e.g. review [6]). However, taking into account the small thickness of the F layer and the small size of domains ($\xi_S \sim$ *10-100nm)*, we estimate the GISANS intensity as being $I_{GISANS} \sim 10^{-6} I_0$ ($I_0$ is intensity of the direct beam) which is below the typical background limit of neutron instruments. In order to enhance the weak intensity of GISANS scattering from thin F layers a waveguide enhancement of neutron density can be used [5]. To do so, the Nb/Gd system should be placed between layers with high neutron scattering length densities (SLDs). The difficulty of neutron studies of systems with gadolinium is the large neutron absorption cross sections for isotopes $^{157}$Gd and $^{155}$Gd which have 16% and 15% concentration in the natural mixture of isotopes. The aim of this article is the design of neutron waveguides with Nb/Gd taking into account all the features of Gd interaction with neutrons.

## 2. Samples design

For optimization procedure a S/F bilayer Nb($d_S$)/Gd($d_F$) and S/F/S Nb($d_{S1}$)/Gd($d_F$)/Nb($d_{S2}$) trilayer systems were considered. While S/F bilayers are simpler to prepare in the technological sense, trilayers S/F/S, due to the presence of superconductors on both interfaces of the F layer may provide more Cooper pair inside it. The thickness of the superconductor must be chosen to be large enough to create superconductivity with $T_C$ close to its bulk value. Following to works [4],[5] we have chosen $d_S$ = 50nm for the S/F bilayer and $d_{S1} = d_{S2}$ = 25nm for the S/F/S trilayer. To ensure penetration of Cooper pairs in the whole F layer the Gd layer thickness must be chosen to be of the order of the superconducting correlation length $\xi_F$ in the F layer ($\xi_F \approx$ 1.4nm for Gd, see [4]). On the other hand $d_F$ must be thick enough to guarantee homogeneous deposition of the F layer. Taking into account these considerations we have fixed $d_F$ = 2nm. Next, the investigated bilayer Gd(2nm)/Nb(50nm) and trilayer Nb(25nm)/Gd(2nm)/Nb(25nm) have to be placed between layers with high SLDs to create a waveguide structure. As the substrate we have chosen sapphire since it has high SLD ($\rho = 5.7 \times 10^{-4}$ nm$^{-2}$) and allows growing structures with small roughness [9]. As the second layer copper can be used with $\rho = 6.6 \times 10^{-4}$ nm$^{-2}$. To protect the copper layer against oxidation a capping layer of Ta(3nm) was considered in calculations.

The presence of the waveguide enhancement in systems with absorbing and ferromagnetic Gd layer can be seen either via dips on the total reflection plateau of non-spin-flip (NSF) reflectivities [10],[11] or as peaks in the spin-flip (SF) channel [7]. In the latter case the vector of magnetization should lie in the sample plane and be normal to the external magnetic field. For our calculations we assumed the Gd magnetization, being equal to its bulk value of 20kG, to be turned normally to the external magnetic field (see insets to Figure 1). To describe the high absorption of Gd, the non-zero imaginary part of the neutron coherent scattering length should be taken into account. We used value $b_{Gd}$ = 6.0-11i fm [13] and bulk density $N_{Gd}$ = 30 nm$^{-3}$ which gives us the scattering length density of Gd $\rho$ = (1.8 -3.3i)$\times 10^{-4}$ nm$^{-2}$.

To optimize the structures we have calculated the minimum of NSF reflectivity $R^{++}$ and maximum of SF reflectivity $R^{+-}$ varying the thickness of the Cu layer $d_1$ (Figure 1). The thickness dependence of the min($R^{++}$) and max($R^{+-}$) for the structure Ta(3nm)/Cu($d_1$)/Gd(2nm)/Nb(50nm)/Al$_2$O$_3$ is shown in Figure.1a. One can see that both min($R^{++}$) and max($R^{+-}$) have the best values at $d_1$ = 17nm. The thickness of the upper layer of the order of tens nanometers is typical for such waveguide structures [7],[8],[14]. In contrast, for S/F/S structure Ta(3nm)/Cu($d_1$)/Nb(25nm)/Gd(2nm)/Nb(25nm)/Al$_2$O$_3$ the effect is maximized at a much smaller value $d_1$ = 4nm. The reason for this lies in the large neutron capture cross-section for gadolinium. The effect is somehow similar to the well-known in nuclear reactor technique the *reactor poisoning* effect, when the presence of highly absorbing isotopes in the active zone of a reactor prevents the formation of neutron density sufficient for reactor operation. In our case the neutron capture by Gd atoms prevents appearing enhanced by the waveguide structure neutron density. To check this we calculated the dependence of max($R^{+-}$) vs $d_1$ for the structure with non-absorbing Gd (see dashed line in Figure 1b). Indeed, one can see that for this case a stronger enhancement can be observed for much thicker Cu layers.

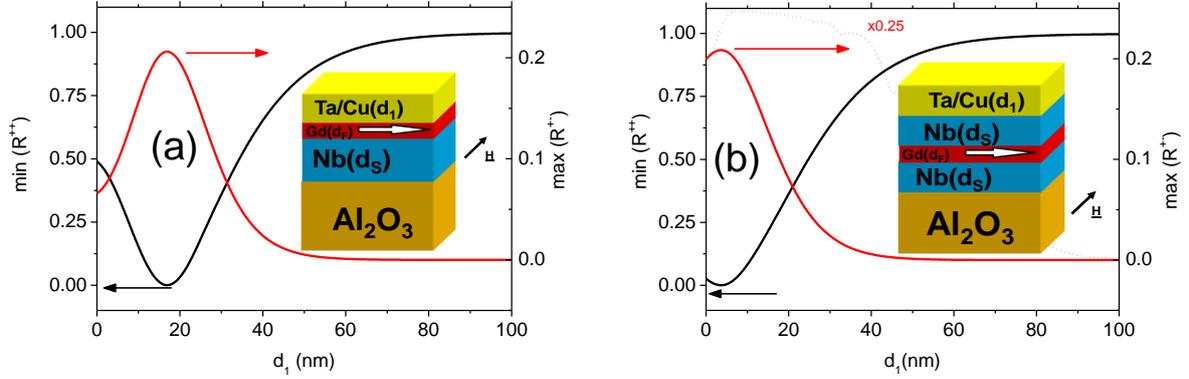

**Figure 1**. (a) Dependence of the dip on the total reflection plateau for non-spin-flip $R^{++}$ curve (black) and peak maximum for spin-flip $R^{+-}$ curve (red) for the S/F bilayer Ta(3nm)/Cu($d_1$)/Gd(2nm)/Nb(50nm)/Al$_2$O$_3$ and (b) S/F/S trilayer Ta(3nm)/Cu($d_1$)/Nb(25nm)/Gd(2nm)/Nb(25nm)/Al$_2$O$_3$. Dashed line is max($R^{+-}$) for the same structure but without imaginary part of SLD in Gd. Note that this line is divided by 4. Insets show sketches of the structures.

## 3. Sample preparation and experimental results

The samples of nominal structure Ta(3nm)/Cu(17nm)/Gd(1-3nm)/Nb(50nm) and Ta(3nm)/Cu(4nm)/Nb(25nm)/Gd(1-3nm)/Nb(25nm) were prepared using an UHV magnetron machine ULVAC MPS-4000-C6 at constant current onto Al$_2$O$_3$(1-102) substrates. Before the deposition the substrates were cleaned from organic contaminations with acetone and alcohol. The substrates were further cleaned with reverse magnetron sputtering (2 minutes at argon flow rate of 25 sccm) in the load chamber. The base pressure was lower than $2\times10^{-9}$ mbar. Pure argon gas (99.9998% purity) at flow rate of 25 sccm was used as sputter gas. The deposition was carried out at room temperature (about 25 °C) at magnetron sputtering power of 100 Watts in argon atmosphere of $1\times10^{-3}$ mbar. In these conditions Nb, Gd, Cu, and Ta layers were sputtered at deposition rates of 2.35 nm/min, 6.85 nm/min, 6.45nm/min, and 2.8 nm/min, respectively. The deposition rates were calibrated using test samples with the help of Zygo NewView7300 white light interferometer.

The PNR experiment was conducted on the angle-dispersive reflectometer NREX at the research reactor FRM-II in Garching, Germany. A polarized neutron beam with wavelength 4.26 ± 0.06 Å and polarization 99.99% falls on the sample under grazing incidence angles $\theta_1$ = [0.25 - 4]°. The divergence of the beam was set to $\Delta\theta_1$ = 0.025° by two slits before the sample. A position sensitive detector was used to measure for every $\theta_1$ intensity of scattered neutrons for a range of outgoing angles $\theta_2$. By switching the spin-flipper state placed before the sample spin-up and spin-down reflectivities were measured. Two samples Ta(3nm)/Cu(17nm)/Gd(3nm)/Nb(50nm) (Gd/Nb) and Ta(3nm)/Cu(4nm)/Nb(25nm)/Gd(3nm)/Nb(25nm) (Nb/Gd/Nb) were measured. Before the measurement the Gd/Nb (Nb/Gd/Nb) sample was cooled down to $T$ = 9K (6K) in magnetic field $H$ = 4.5kG applied parallel to the surface and normal to the scattering plane. After the cooling down the magnetic field was set to $H$ = 3kG (4kG). Temperatures and magnetic fields were chosen to be above $T_C$ of the S layer and the saturation field of the F layer, which were determined in advance with SQUID magnetometry. Both samples have shown difference between the spin-up $R^+$ and spin-down $R^-$ reflectivities which gives the direct evidence of the presence of ferromagnetism in the Gd layer.

However, since magnetic and superconducting properties are out of the scope of this article, we will discuss them elsewhere.

Figures 2a,b show 2D scattering maps for spin-down neutrons for the Gd/Nb and Nb/Gd/Nb samples correspondingly. Tilted line at the condition $\theta_2 = \theta_1$ corresponds to the specular reflection. The presence of the waveguide enhancement can be seen by the dip on the total reflection plateau at the angle $\theta_{res} = 0.3°$ for both Gd/Nb and Nb/Gd/Nb samples. This dip can be more clearly seen in the insets to Figures 2b,c where the reflectivity curves are presented. Apart of the dip, the resonant enhancement is proved by the presence of channeled beam ($\theta_2 \approx 0$) [12].

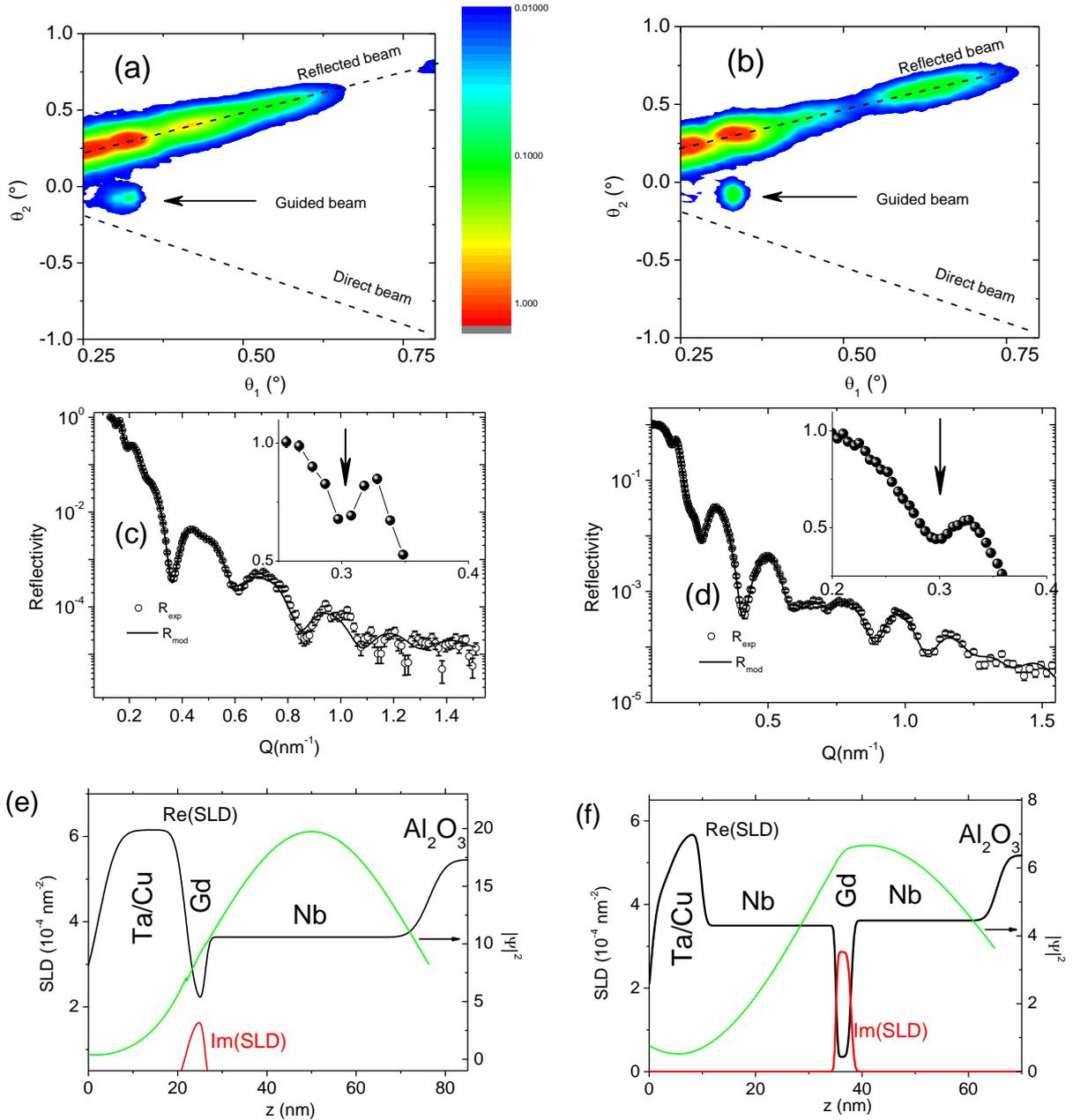

**Figure 2.** Scattering maps for the Gd/Nb (a) and Nb/Gd/Nb (b) samples. Experimental (dots) and model (solid lines) reflectivity curves for the Gd/Nb (a) and Nb/Gd/Nb (b) samples. Insets show the experimental curves around the resonance (shown by arrows) as a function of $\theta_1$. The depth profiles of SLD and neutron density for the Gd/Nb sample (e) and Nb/Gd/Nb sample (f).

## 4. Discussions and conclusion

Specular reflectivities were fit to the model curves. Best-fit curves are shown in Figure 2c,d by solid lines. The resulting depth profiles of SLD are shown in Figures 2e,f. From the fit it follows that the real thicknesses of the Nb and Gd layers are close (within 10% maximum) to their nominal values. The fit also shows a rather small rms roughness of the Gd/Nb interface ($\sigma = 0.7$nm for the Gd/Nb and $\sigma = 0.4$nm for the Nb/Gd/Nb sample). The imaginary part of SLD for gadolinium in Gd/Nb and Nb/Gd/Nb samples are 50% and 88% of bulk value, respectively.

Having the SLD depth profile known we can calculate the depth profiles of the neutron density in the waveguide mode (Figure 2e,f). One can see that the neutron density in both cases has a maximum in the center of Nb layer and makes value $|\Psi|^2 = 20$ and $|\Psi|^2 = 7$ for the Gd/Nb and Nb/Gd/Nb samples. This small enhancement is explained by the *poisoning effect* as described in Section 2. However, despite of this effect, the neutron density around the Gd layer in both cases is similar for both systems ($|\Psi|^2 \approx 7$-9). Such an enhancement will allow GISANS intensity from magnetic domains to be increased up to a detectable level of $I_{GISANS} \sim 10^{-5}$-$10^{-4} I_0$.

In conclusion, the features of the formation of the enhanced standing wave in the structure with a thin highly absorbing layer of gadolinium are considered in the article. The analogue of the *poisoning effect* well known in reactor physic was found. Effect is higher when Gd layer is placed in the middle of the waveguide. Despite of this effect for Nb/Gd bilayer and Nb/Gd/Nb trilayer placed between the $Al_2O_3$ substrate and Cu layer it was shown theoretically and experimentally that one order of magnitude enhancement of neutron density is possible in the vicinity of Gd layer. This enhancement makes it possible to study domain formation in Gd layer under transition of Nb layer(s) into superconducting state (cryptoferromagnetic phase).

The authors would like to thank A.I. Frank, G. Kulin and S. Kozhevnikov for fruitful discussion. This work is based upon experiments performed at the NREX instrument operated by Max-Planck Society at the Heinz Maier-Leibnitz Zentrum (MLZ), Garching, Germany and partially supported by DFG collaborative research center TRR 80 and RFBR (Projects 14-22-01007, 14-22-01063).